\documentclass[12pt]{article}

\usepackage{amsmath}
\usepackage{amsfonts}
\usepackage{amssymb}
\usepackage{graphicx}
\usepackage{psfrag}
\usepackage{graphicx}
\usepackage{psfrag}

\newcommand{\be}{\begin{equation}}
\newcommand{\ee}{\end{equation}}
\newcommand{\ba}{\begin{eqnarray}}
\newcommand{\ea}{\end{eqnarray}}

\begin{document}
\begin{center}
{\bf
SOME PROPERTIES OF THE SHAPE INVARIANT TWO-DIMENSIONAL SCARF~II MODEL}\\
\vspace{1cm}
{\large \bf
M. V. Iof\/fe$^{1,}$\footnote{E-mail: m.ioffe@spbu.ru},
E. V. Kolevatova$^{1,}$\footnote{E-mail: e.v.krup@yandex.ru},\\
D. N. Nishnianidze}$^{2,1,}$\footnote{E-mail: cutaisi@yahoo.com}\\
\vspace{0.5cm}
$^1$ Saint Petersburg State University, 198504 Saint-Petersburg, Russia.\\
$^2$ Akaki Tsereteli State University, 4600 Kutaisi, Georgia.\\
\vspace{1.0cm}

\end{center}
\vspace{0.5cm}
\hspace*{0.5in}
\vspace{1cm}
\hspace*{0.5in}
\begin{minipage}{5.0in}
{\small
Two-dimensional Scarf~II quantum model is considered in the framework of Supersymmetrical
Quantum Mechanics (SUSY QM). Previously obtained results for this integrable system are systematized, and some new properties are derived. In particular, it is shown that
the model is exactly or quasi-exactly solvable in different regions of parameter of the system. The degeneracy of the spectrum is
detected for some specific values of parameters. The action of the symmetry operators of fourth order in momenta is calculated
for the arbitrary wave functions, obtained by means of double shape invariance.
}
\end{minipage}



\section*{\normalsize\bf 1. \quad Introduction.}
\vspace*{0.5cm}
\hspace*{3ex} Two different generalizations of the supersymmetrical approach \cite{review} in Quantum Mechanics from one-dimensional
to two-dimensional case are known. The first one deals with the supercharges linear in momenta, and the corresponding SuperHamiltonian includes
both a pair of scalar components and one matrix component \cite{abei}. The matrix component can be interpreted, for example, as a Pauli Hamiltonian
for nonrelativistic fermions moving in the plane with external electromagnetic field \cite{pauli}. As opposed to this, an alternative supersymmetrical approach \cite{david} with supercharges of second order in momenta avoids any matrix component in the SuperHamiltonian. In this method, the main initial step is the solution of supersymmetrical intertwining relations between a pair of scalar Schr\"odinger Hamiltonians. Intertwined operators - second order supercharges - were classified according to the metrics in highest derivatives \cite{david}. For the simplest case of supercharges with Lorentz (hyperbolic) metrics, all possible solutions for potentials were found explicitly \cite{general}.

In one-dimensional case, among the superpartner scalar Hamiltonians the particular class with the so called shape invariance property exists \cite{shape}, \cite{review}, \cite{mallow}. Usually, it provides an elegant purely algebraic method to solve the problem, i.e. to find analytically both the energy spectrum and corresponding wave functions. All well known exactly solvable one-dimensional Hamiltonians belong \cite{dabrowska} to
the class of shape invariant ones (some additional unknown exactly solvable systems were built \cite{2014} just by the SUSY method). The notion of shape invariance was generalized to the case of two-dimensional second order Supersymmetrical Quantum Mechanics \cite{newmethod}, \cite{shape-2}, \cite{ioffe1}, \cite{ioffe2} as well. In this situation, a'priori the shape invariance does not guarantee the solvability of the system, but for some models it helps, together with other tools, to solve \cite{newmethod}, \cite{morse}, \cite{poschl}, \cite{scarf} the problem exactly or quasi-exactly \cite{quasi}.

There are some details in two-dimensional generalizations of SUSY Quantum Mechanics, which have to be investigated to make the formulation complete. We mean the relationship between exactly solvable and quasi-exactly solvable models with the same potential, the action of the symmetry operators onto the built wave functions, especially in the case of possible degeneracy of the spectra. As we shall demonstrate, the latter is realized for some specific values of coupling constants. Just this range of problems will be studied in the present paper in the case of Scarf~II model - chronologically the last constructed solvable model, which allows to apply all tools mentioned above.

The paper is organized as follows. Section 2 contains formulation of two-dimensional Scarf~II system, and two different forms of special SUSY separation of variables are represented. One of them (Subsection 2.1) provides exact solvability of the model for discrete negative values $a_k=-k$ of parameter $a.$ It is shown in Subsection 2.2, that another method of SUSY separation works for parameter $a > -1/2$ which does not intersect with the previous one. In this case, a part of wave functions is obtained by means of intertwining relations and shape invariance.
The details concerning possible degeneracy of energy levels and action of symmetry operators are considered in Section 3. It is shown that all constructed wave functions are simultaneously eigenfunctions of symmetry operators.

\section*{\normalsize\bf 2. \quad Scarf~II Model: exact and quasi-exact solvability.}
\vspace*{0.5cm}
\hspace*{3ex} The intertwining relations:
\begin{equation}\label{int}
H^{(1)} Q^+ = Q^+ H^{(2)};\quad Q^- H^{(1)} = H^{(2)} Q^-
\end{equation}
between a pair of Hamiltonians
\begin{equation}\label{ham}
H^{(i)}(\vec x; a) = - \Delta^{(2)}+V^{(i)}(\vec x; a);\,\, i=1,2;\quad \vec x = (x_1, x_2);\quad \Delta^{(2)}\equiv \partial_1^2+\partial_2^2;\quad
\partial_i\equiv \frac{\partial}{\partial x_i}
\end{equation}
play the fundamental role in the supersymmetrical approach in Quantum Mechanics. In the case of Lorentz metrics in supercharges, one of solutions \cite{shape-2} for $Q^{\pm}$ is:
\ba
&&Q^{+}=(Q^{-})^{\dag}=4\partial_{+}\partial_{-}+2a\tanh (x_{+}/2)\partial_{-}+2a\coth (x_{-}/2)\partial_{+}+a^{2}\tanh (x_{+}/2)
\coth (x_{-}/2)-\label{Q}\\
&&-(\cosh x_{1})^{-2}\biggl[c(2b+1)\sinh x_{1}+(c^2-b^2-b)\biggr]+(\cosh x_{2})^{-2}\biggl[c(2b+1)\sinh x_{2}+(c^2-b^2-b)\biggr] , \nonumber
\ea
with $x_{\pm}\equiv x_{1} \pm x_{2}, \partial_{\pm}=\partial / \partial x_{\pm},$ and real parameters $a$ and $b, c >0,$ while the corresponding superpartner Hamiltonians are the two-dimensional generalizations of Scarf~II model:
\be
H^{(i)}(\vec x; a) = -\Delta^{(2)}-\frac{a(a\mp 1)}{2}\bigl(\frac{1}{\cosh^{2}(x_{+}/2)}-\frac{1}{\sinh^{2}(x_{-}/2)}\bigr) + U(x_{1}) + U(x_{2}), \label{hamm}
\ee
where one-dimensional potential $U$ has just the form of known Scarf~II potential:
\be
U(x)= \frac{c(2b+1)\sinh x+(c^2-b^2-b)}{\cosh^{2}x}. \label{U}
\ee
Here and below, for simplicity we skip sometimes the explicit dependence on some of parameters, but we'll restore this dependency as needed.

\subsection*{\normalsize\bf 2.1. \quad Exact solvability.}
\vspace*{0.5cm}
\hspace*{3ex}
Both Hamiltonians $H^{(i)}$ include the mixing term which prevents the standard procedure of separation of variables. But the structure of the mixing term allows
to apply the special procedure - SUSY separation of variables \cite{ioffe2}. Namely, choosing \cite{scarf1} the specific value of parameter $a=-1$ we obtain one of the superpartner
Hamiltonians, $H^{(2)},$ amenable to the standard separation. More of that, one-dimensional Schr\"odinger equations with Scarf~II potentials (\ref{U}) are exactly solvable \cite{scarf}. These two circumstances - separation of variables and solvability of one-dimensional problems - allow to solve the Schr\"odinger equation with Hamiltonian $H^{(2)}$ completely, i.e. to find analytically all its energy eigenvalues and its wave functions as usually in problems with separation of variables:
\be
\Psi^{(2)\pm}_{n,m}(\vec x)=\eta_n(x_1)\eta_m(x_2) \pm \eta_m(x_1)\eta_n(x_2),
\label{sep}
\ee
where $\eta (x)$ - known solutions of one-dimensional Scarf II problem:
\be
\eta_n(x)=(\cosh x)^{-b}\exp[-c\cdot \arctan(\sinh x)]P_{n}^{(\gamma, \beta)}(i\sinh x);\quad
\gamma \equiv  -(b + ic + 1/2);\, \beta \equiv \gamma^{\star}, \label{eta}
\ee
expressed in terms of Jacobi polynomials \cite{jacobi}.

In turn, the Hamiltonian $H^{(1)}$ still contains mixing term, and its spectr can be investigated by means of the intertwining relations (\ref{int}). Namely, all its normalizable wave functions can be obtained as:
\be
\Psi^{(1)}_{n,m}(\vec x) = Q^+\Psi_{n,m}^{(2)-}
\label{an}
\ee
being expressed in terms of Jacobi polynomials (see Eqs.(9), (11), (12) in \cite{scarf1}). The symmetric functions $\Psi^{(2)+}$ were excluded in (\ref{an}) since they would lead to non-normalizable $\Psi^{(1)}.$ The discrete energy bound state values are:
\be
\varepsilon^{(1)}_{n,m}(a=-1) = -(b-n)^2 - (b-m)^2; \quad n,m < b;\quad |n-m| > 1.
\label{energy}
\ee

The result above was obtained for the parameter value $a=-1,$ and it can be extended to the arbitrary negative integer $a_k=-k$ by means of
the shape invariance property: superpartner Hamiltonians $H^{(1)},\, H^{(2)}$ in (\ref{hamm}) are related as follows:
\be
H^{(1)}(a_k) = H^{(2)}(a_{k+1}),
\label{shape}
\ee
where dependence on parameter $a$ is pointed out explicitly. For these values of $a,$ the discrete energy
spectrum $\varepsilon_{n,m}(a=-k)$ of $H^{(1)}(a=-k)$ is expressed by the same formula (\ref{energy}), but now with $|n-m| > k,$ only. The seeming degeneracy of obtained energy levels under replacing $n \rightleftarrows m$ is absent since, due to singular properties of $Q^+,$ only the symmetric eigenfunctions $\Psi^{(1)}$ are normalizable. The corresponding analytical
expressions for the wave functions $\Psi^{(1)}(\vec x; a=-k)$ can be derived \cite{scarf1} analytically by means of (\ref{an}), they depend explicitly on parameter $a_k=-k.$

\subsection*{\normalsize\bf 2.2. \quad Quasi-exact solvability.}
\vspace*{0.5cm}
\hspace*{3ex}
It is interesting to compare the above results with those obtained by another method - the Second SUSY separation procedure \cite{ioffe2} -
for the same Hamiltonian $H^{(1)}$ from (\ref{hamm}). This method works due to separation of variables for the intertwining supercharge $Q^-.$
It was shown \cite{david}, \cite{ioffe1} that such separation of variables in $Q^-$ is possible in the case of Lorentz metrics for arbitrary coefficient functions due to a suitable
gauge (nonunitary) transformation of the problem. In particular, this is true for the specific coefficients appeared in (\ref{Q}). For the first time, this
procedure was proposed for the two-dimensional Morse potential in \cite{newmethod}, it was explored again for the two-dimensional P\"oschl-Teller potential in
\cite{poschl}, and finally, for two-dimensional Scarf~II model - in \cite{scarf2}. We shall reproduce very briefly the main steps of the latter but adopting them for
the slightly different form of potential. This is necessary to compare results with those described above for $a_k=-k$ values.

We start again from the same Hamiltonians (\ref{hamm}), (\ref{U}) satisfying intertwining relation (\ref{int}) with the conjugated supercharge $Q^-$ given
by (\ref{Q}).
It is useful to perform the following similarity (gauge) transformation:
\be
Q^-=\exp{(-\chi )} q^- \exp{(+\chi )};\quad \chi(\vec x)=-a\ln|\cosh(x_+/2)\sinh(x_-/2)|,\label{gauge}
\ee
in order to separate variables $x_1,\,x_2$ in supercharge:
\be
q^-= \partial_1^2-\partial_2^2 - U(x_1) + U(x_2). \label{q-}
\ee
Fortunately, like in Subsection 2.1 the explicit form of one-dimensional "potentials" $U(x)$ in (\ref{q-}) after separation of variables is amenable to analytic solution,
very rare situation. The zero modes $q^-\omega_n(\vec x)=0$ of auxiliary operator $q^-$ can be written as a sum of products:
\be
\omega_n(\vec x)=\eta_n(x_1)\eta_n(x_2);\quad \biggl(-\partial_1^2 + U(x)\biggr)\eta_n(x) = \epsilon_n\eta_n(x); \quad \epsilon_n=-(b-n)^2;\quad n<b. \label{etaa}
\ee
The solutions of Schr\"odinger-like equation (\ref{etaa}) are known as solutions (\ref{eta}) of one-dimensional Scarf II problem being expressed analytically
in terms of Jacobi polynomials. Taking into account the similarity transformation (\ref{gauge}), one obtains the zero modes $\Omega_n(\vec x)$
of initial supercharge $Q^-:$
\ba
\Omega_n(\vec x)=
\exp{(-\chi(\vec x))}\omega_n(\vec x)=|\cosh(x_+/2)\sinh(x_-/2)|^{a}\eta_n(x_1)\eta_n(x_2). \label{P}
\ea
The so called "second SUSY separation of variables procedure" is based \cite{newmethod}, \cite{ioffe1}, \cite{ioffe2} on a construction of {\it normalizable} zero modes $\Omega_n,$ whose proper linear
combinations will provide some subset (quasi-exact solvability !) of normalized wave functions of $H^{(1)}.$ It is clear from expressions (\ref{P}) that this is possible for arbitrary values of parameter $a$ but with the only restriction $a > -1/2 :$ otherwise, $\Omega_n$ being decreasing at infinity is not $L_2-$integrable along the line $x_-=0.$ Therefore, wave functions which can be built by means of the algorithm of second SUSY separation of variables correspond to the values of parameter $a$, not intersecting
with values in the first procedure of SUSY separation where parameter $a$ took discrete values $a_k=-k,\, k=1,2,...$.

Due to second of intertwining relations (\ref{int}), the variety of zero modes $\Omega_n$ is closed under the action of operator $H^{(1)},$ i.e. the action of $H^{(1)}$ onto $\Omega_n$ equal to the linear combination of $\Omega'$s:
\be
H^{(1)}\Omega_n(\vec x)=\sum_{k=0}^{N}c_{nk}\Omega_k(\vec x),
\label{C}
\ee
where $c_{nk}$ form the $(N+1)\times (N+1)$ constant matrix $\widehat C.$ One can check that matrix $\widehat C$ is triangular, and therefore its eigenvalues coincide with diagonal elements which in turn give the discrete energy spectrum $E_n$ of the Hamiltonian $H^{(1)}.$ By default the parameter $a$ value is implied here fixed in the region $a > -1/2 .$ The result of calculations of diagonal elements of $\widehat C$ is:
\be
c_{n,n}=E_{n,0}=2\epsilon_n - a^2 + 2a(b-n)=-(b-n)^2-(b-a-n)^2;\quad a>-1/2.
\label{diag}
\ee
As for the wave functions $\Psi_n^{(1)}(\vec x; a),$ they can be obtained analytically by means of explicit diagonalization of matrix $\widehat C,$ which is performed by the
recurrent  algorithm \cite{newmethod}, \cite{ioffe1}, \cite{ioffe2}, \cite{scarf2}. Thus, for these values of parameter $a$ we deal with quasi-exactly-solvable class of models: we know a part of spectrum and their wave functions.

The shape invariance property of considered model:
\be
H^{(2)}(\vec x; a) = H^{(1)}(\vec x; a+1)
\label{sh}
\ee
allows to extend the variety of known wave functions and corresponding energies for this quasi-exactly-solvable models. Namely, for each - "principal" - wave function
$\Psi^{(1)}_{n,0}(\vec x; a+m)$ found above (but with the parameter value $(a+m)$ instead of $a$) one can build the whole set of new wave
functions $\Psi^{(1)}_{n,m}(\vec x; a)$ with parameter $a$ as:
\be
\Psi^{(1)}_{n,m}(\vec x; a)=Q^+(a)Q^+(a+1)...Q^+(a+m-1)\Psi^{(1)}_{n,0}(\vec x; a+m),
\label{chain1}
\ee
with corresponding energies:
\be
E^{(1)}_{n,m}(a)=E^{(1)}_{n,0}(a+m)=-(b-n)^2-(b-a-m-n)^2.
\label{en}
\ee
It is necessary to make a remark concerning notations. In Eq.(\ref{en}), the first index $n$ of energy value numerates the ordinal number of diagonal element
of matrix $\widehat C,$ i.e. the ordinal number of "principal" wave function built as a linear combination of zero modes $\Omega ,$ while the second index $m$
numerates the number of operators $Q^+$ in a chain (\ref{chain1}).

The considered quasi-exactly-solvable model obeys one more symmetry property which allows to expand further the class of wave functions known analytically.
We mean the additional - second - shape invariance of the system \cite{scarf2}. Let us remind that the Hamiltonian $H^{(1)}(\vec x)$
from (\ref{hamm}) depends not only on parameter (coupling constant) $a,$ but also on two others -- $b$ and $c.$ Briefly reformulating observations noticed in \cite{shape-2}, \cite{scarf2}, the Hamiltonian $H^{(1)}(\vec x)$ participates also in a different pair of intertwining relations. It is convenient to accept the identity $H^{(1)}\equiv \widetilde H^{(1)}.$ Then:
\be
\widetilde H^{(1)} \widetilde Q^+ = \widetilde Q^+ \widetilde H^{(2)};\quad \widetilde Q^- \widetilde H^{(1)} = \widetilde H^{(2)} \widetilde Q^-.
\label{int2}
\ee
These intertwining relations have the same form as (\ref{int}), the first Hamiltonian is the same, but the difference is in specific expressions for operators $\widetilde Q^{\pm},\, \widetilde H^{(2)}.$
Namely, the superpartner Hamiltonians in (\ref{int2}) are:
\ba
&&\widetilde{H}^{(1),(2)}(\vec x)=-\Delta^{(2)}-\frac{a(a-1)}{2}(\frac{1}{\cosh^{2}(x_{+}/2)}-\frac{1}{\sinh^{2}(x_{-}/2)})+
\label{3.2}\\
&&+\frac{c(2b\pm 1)\sinh x_{1}-b(b\pm 1)+c^2}{\cosh^{2}x_{1}}
+\frac{c(2b\pm 1)\sinh x_{2}-b(b\pm 1)+c^2}{\cosh^{2}x_{2}},\nonumber
\ea
and they are intertwined by means of new supercharges:
\ba
&&\tilde{Q}^\pm =\partial_1\partial_2\mp(b\tanh x_1+\frac{c}{\cosh x_1})\partial_2\mp(b\tanh x_2+\frac{c}{\cosh x_2})\partial_1+
\nonumber\\&&+
(b\tanh x_1+\frac{c}{\cosh x_1})(b\tanh x_2+\frac{c}{\cosh x_2})+\frac{a(a-1)}{4\cosh^2(x_+/2)}+\frac{a(a-1)}{4\sinh^2(x_-/2)}.
\label{Qnew}
\ea
It is important that new Hamiltonian $\widetilde H^{(2)}$ does not coincide with the initial $H^{(2)}$.
One can easily check that the pair $\widetilde H^{(1)}\equiv H^{(1)}$ and $\widetilde H^{(2)}$ also obeys the property of shape invariance but under parameter $b$ now:
\be
H^{(1)}(\vec x; a, b) = \widetilde H^{(2)}(\vec x; a, b+1).
\label{sh2}
\ee

Analogously to the previous procedure, the intertwining (\ref{int2}) allows to build new wave functions $\widetilde\Psi^{(1)}$ starting from an arbitrary "principal" wave function. The role of this "principal" state can be played by the wave functions $\Psi^{(1)}_{n,m}(\vec x; a, b)$ obtained above in (\ref{chain1}). Thus, we have a variety of wave functions of the Hamiltonian $H^{(1)}(\vec x; a, b)$ which are written as a product of some number of operators $\widetilde Q^+$ and some number of operators $Q^+$ onto "principal" function $\Psi^{(1)}_{n,0}(\vec x):$
\ba
&&\Psi^{(1)}_{n,m,M}(\vec x; a, b)= \widetilde Q^+(a, b)\widetilde Q^+(a, b-1) ... \widetilde Q^+(a, b-M+1)\Psi^{(1)}_{n,m}(\vec x; a, b-M)=\nonumber\\
&&=\widetilde Q^+(a, b)\widetilde Q^+(a, b-1)...\widetilde Q^+(a, b-M+1)\cdot\nonumber\\
&&\cdot Q^+(a, b-M)Q^+(a+1, b-M)...Q^+(a+m-1, b-M)\Psi^{(1)}_{n,0}(\vec x; a+m, b-M),
\label{chain2}
\ea
where new index $M$ marks a number of operators $\widetilde Q^+.$ The positions of operators $\widetilde Q^+$ and $Q^+$ in this chain can be changed by means of relation:
\ba
\widetilde Q^{+}(\vec x; a, b)Q^{+}(\vec x; a, b-1)=Q^{+}(\vec x; a, b)\widetilde Q^{+}(\vec x; a+1, b) \label{3.9}
\ea
both its sides intertwine the same pair of Hamiltonians: $H^{(1)}(\vec x; a ,b)$ and $H^{(1)}(\vec x; a+1 ,b-1).$

\section*{\normalsize\bf 3. \quad Integrability of the system.}
\vspace*{0.5cm}
\hspace*{3ex} One can check that the system with Hamiltonian $H^{(1)},$ which satisfies the intertwining relations, obeys \cite{david} the symmetry operator of fourth order in derivatives $R^{(1)},$ and the explicit form of operator $R^{(1)}$ is known in terms of supercharges:
\be
R^{(1)}=Q^+Q^-;\quad [H^{(1)}, \, R^{(1)}]=0.
\label{sym}
\ee
This symmetry property of Hamiltonian is a general consequence of intertwining relations: an arbitrary Hermitian operator which is intertwined with "anything" by operators
$Q^{\pm},$ commutes with corresponding symmetry operator $R^{(1)}.$ Meanwhile, it is known \cite{miller} that for any two-dimensional system the number of mutually commuting symmetry operators  can not be two or more, i.e. $H^{(1)}$ describes a completely integrable model.

The spectrum (\ref{en}) found above is not degenerate, and therefore, all wave functions (\ref{chain1}) of $H^{(1)}$ must be eigenfunctions of $R^{(1)},$ as well.
This property is not obvious from the explicit form of (\ref{chain1}) with arbitrary number of multipliers $Q^+(a+k).$ In this sense, it would be interesting to check this property directly. As a motivating remark, in the case of two-dimensional Morse potential \cite{newmethod}, three additional wave functions, besides that of the form analogous to (\ref{chain1}), were constructed "by hands". For each of them, it was demonstrated by direct calculation that it is an eigenfunction of the symmetry operator. Also, in our present case, an accidental degeneracy of the spectrum is possible for some values of coupling constants. Namely, according to (\ref{en}) two energy levels $E_{n,m}$ and $E_{n',m'}$ of wave functions (\ref{chain1}) may coincide for the model with specific relation between two parameters $a$ and $b$ of the model:
\ba
&&a(\delta m-\delta n) + b(2\delta n+\delta m)=(n+n')\delta n +1/2 (m+m')\delta m + (nm-n'm');\label{degen}\\
&&\delta n\equiv n-n';\, \delta m\equiv m-m'.
\nonumber
\ea
In general, if this relation is fulfilled, action of the symmetry operator $R^{(1)}$ onto $\Psi^{(1)}_{n,m}$ may transform it into another wave function $\Psi^{(1)}_{n',m'}$ with the same energy. Of course, this problem exists also for more general wave functions obtained by a product of $Q^+$ and $\widetilde Q^+.$

Thus, it is necessary to calculate how the symmetry operator acts onto an arbitrary wave function of this form. We shall prove by the standard method of mathematical induction that an arbitrary eigenfunction (\ref{chain2}) of $H^{(1)}$ is simultaneously an eigenfunction of the symmetry operator $R^{(1)}.$ For simplicity, we shall start from the functions of the form (\ref{chain1}). First of all, the useful relation exists between $R^{(1)}(a)$ and $R^{(2)}(a-1).$ Due to shape invariance (\ref{sh}), both operators $R^{(1)}(a)$ and $R^{(2)}(a-1)$ are the symmetry operators for the same Hamiltonian $H^{(1)}(a),$ and therefore, they must coincide up to the function of the $H^{(1)}(a)$ itself. It is obvious that actually the only opportunity is:
\be
R^{(1)}(a)=R^{(2)}(a-1)+\alpha (a)H^{(1)}(a)+\beta (a),
\label{RR}
\ee
where dependence $R$ and $H$ on $\vec x$ is skipped, and $\alpha (a), \beta (a)$ do not depend on $\vec x.$ Explicitly, coefficients $\alpha , \beta$ were calculated in \cite{shape-2} (see equation after Eq.(34)):
\be
\alpha (a) = 2(2a-1);\quad \beta (a) = (2a-1)(2a^2-2a-3).
\ee
The relation analogous to (\ref{RR}) exists between the operators $\widetilde R^{(1)}(a, b)$ and $\widetilde R^{(2)}(a, b+1).$

Now, to start induction ($b-$dependence is skipped):
\ba
&&R^{(1)}(a)\Psi_{n,1}^{(1)}(\vec x; a)=Q^+(a)Q^-(a)Q^+(a)\Psi_{n,0}^{(1)}(\vec x; a+1)=\nonumber\\
&&=Q^+(a)\biggl(Q^+(a+1)Q^-(a+1)-\alpha (a+1)E^{(1)}_{n,0}(a+1)-\beta (a+1)\biggr)\Psi_{n,0}^{(1)}(\vec x; a+1)=\nonumber\\
&&=-\biggl(E^{(1)}_{n,1}(a)\alpha(a+1)+\beta(a+1)\biggr)\Psi^{(1)}_{n,1}(\vec{x};a), \label{ind1}
\ea
where explicit expression (\ref{en}) was used, proves the statement for $m=1.$ The case of $m=2$ can be also calculated analogously, and
the result provides the hypothesis for higher $m.$  Let us suppose that for general value of $m$ the following relation is true:
\be
R^{(1)}(a)\Psi^{(1)}_{n,m}(\vec{x};a)=-\biggl(\sum_{k=1}^{k=m}[E^{(1)}_{n,m}(a)\alpha(a+k)+\beta(a+k)]\biggr)\Psi^{(1)}_{n,m}(\vec{x};a). \label{ind2}
\ee
To perform the second step of induction,
\ba
&&R^{(1)}(a)\Psi^{(1)}_{n,m+1}(\vec{x};a)=R^{(1)}(a)Q^+(a)\Psi^{(1)}_{n,m}(\vec{x};a+1)=\nonumber\\&&=
Q^+(a)\biggl(R^{(1)}(a+1)-\alpha(a+1)E^{(1)}_{n,m}(a+1)-\beta(a+1)\biggr)\Psi^{(1)}_{n,m}(\vec{x};a+1)=\nonumber\\&&=
-\biggl(\sum_{k=1}^{k=m+1}[E^{(1)}_{n,m}(a+1)\alpha(a+k)+\beta(a+k)]\biggr)\Psi^{(1)}_{n,m+1}(\vec{x};a)=\nonumber\\&&=
-\biggl(\sum_{k=1}^{k=m+1}[E^{(1)}_{n,m+1}(a)\alpha(a+k)+\beta(a+k)]\biggr)\Psi^{(1)}_{n,m+1}(\vec{x};a). \label{ind3}
\ea
Thereby, the statement was proved - wave functions of the form (\ref{chain1}) are simultaneously eigenfunctions of the symmetry operator $R^{(1)},$ independently of possible occasional degeneracy of energy levels.

Analogously, the wave functions (\ref{chain2}) obtained by the chain of operators $\widetilde Q^+$ and $Q^+$ are the eigenfunctions of $R^{(1)},$ as well. In this case, the proof is based on the relation between two different forms of symmetry operators for the same Hamiltonian $H^{(1)}:$
\be
R^{(1)}=Q^+Q^-;\quad \widetilde R^{(1)}=\widetilde Q^+ \widetilde Q^-.
\label{RRR}
\ee
Both operators are of the same - fourth - order in derivatives, and no second - in addition to $R^{(1)}$ - independent symmetry operator may exist for two-dimensional system.
This is possible only if both operators in (\ref{RRR}) coincide up to some second order polynomial of Hamiltonian $H^{(1)}$ with constant coefficients. Since expressions for operators in (\ref{RRR}) are rather cumbersome, the coefficients were defined by means of analytical calculations in MATHEMATICA \cite{wolfram}:
\be
R^{(1)}(a, b)=-4\widetilde R^{(1)}(a, b) + (H^{(1)})^2 + (2 a^2 + 4 b^2)H^{(1)}+(a^4 + 4 b^4).
\label{two}
\ee
In contrast to relation (\ref{RR}), relationship between $\widetilde R^{(1)}$ and $R^{(1)}$ has nothing to do with shape invariance.

For illustration, we consider the action of symmetry operator $\widetilde R^{(1)}(a, b)$ onto $\Psi^{(1)}_{n,m,1}(\vec x; a,b),$ i.e. $M=1$ in (\ref{chain2}):
\ba
&&\tilde{R}^{(1)}(a,b)\Psi^{(1)}_{n,m,1}(\vec{x};a,b)=\tilde{R}^{(1)}(a,b)\tilde{Q}^+(a,b)\Psi^{(1)}_{n,m}(\vec{x};a,b-1)=\nonumber\\&&=
\tilde{Q}^+(a,b)\tilde{R}^{(2)}(a,b)\Psi^{(1)}_{n,m}(\vec{x};a,b-1)=\nonumber\\&&=
\tilde{Q}^+(a,b)\biggl(\tilde{R}^{(1)}(a,b-1)+\tilde{\alpha}(a,b)H^{(1)}(a,b-1)+\tilde{\beta}(a,b)\biggr)\Psi^{(1)}_{n,m}(\vec{x};a,b-1)=\nonumber\\&&=
\biggl(\tilde{\alpha}(a,b)E_{n,m}^{(1)}(a,b-1)+\tilde{\beta}(a,b)\biggr)\Psi^{(1)}_{n,m,1}(\vec{x};a,b)+\nonumber\\&&+\tilde{Q}^+(a,b)\tilde{R}^{(1)}(a,b-1)
\Psi^{(1)}_{n,m}(\vec{x};a,b-1).\label{2}
\ea
The last term is calculated by means of (\ref{two}):
\ba
&&\tilde{Q}^+(a,b)\tilde{R}^{(1)}(a,b-1)\Psi^{(1)}_{n,m}(\vec{x};a,b-1)=\tilde{Q}^+(a,b)\biggl(R^{(1)}(a,b-1)-\nonumber\\
&&-c_1(H^{(1)}(a,b-1))^2-c_2H^{(1)}(a,b-1)-c_3\biggr)\Psi^{(1)}_{n,m}(\vec{x};a,b-1)=\nonumber\\
&&=const\cdot\Psi^{(1)}_{n,m,1}(\vec{x};a,b).\label{4}
\ea
Thus,
\be
\tilde{R}^{(1)}(a,b)\Psi^{(1)}_{n,m,1}(\vec{x};a,b)=const\cdot\Psi^{(1)}_{n,m,1}(\vec{x};a,b).
\ee
The case of arbitrary $M>1$ can be proved analogously.

\section*{\normalsize\bf 4. \quad Conclusions.}
\vspace*{0.5cm}
\hspace*{3ex} The present paper continues study of two-dimensional integrable Scarf~II model, which is not amenable to conventional separation of variables procedure.
In particular, the dependence of spectra on parameters of the model, its possible degeneracy and properties of fourth order symmetry operators were investigated.
The analogous properties of two other - Morse and P\"oschl-Teller - two-dimensional models not amenable to separation can be studied in the same way. For example,
the two-dimensional Morse system \cite{newmethod}, \cite{morse} is exactly solvable for its parameter $a=-(k+1)/2;/, k=0,1,2...$ and it is quasi-exactly-solvable for nonoverlapping region $a>1/4+1/(4\sqrt{2}).$

\section*{\bf Acknowledgments.}

The work was partially supported by the grants of Saint Petersburg State University N 11.38.223.2015 (E.V.K.), N 11.42.1303.2014 (M.V.I.) and by the grant RFBR N 13-01-00136-а (M.V.I. and E.V.K.).

\end{document}